\documentclass[conference]{IEEEtran}

\usepackage{helvet}        
\usepackage{microtype}             
\usepackage{inconsolata}              
\usepackage{textcomp}              

\usepackage[numbers,square,sort&compress]{natbib}
\usepackage{flushend}

\usepackage{graphicx}              

\usepackage{booktabs}              

\usepackage{relsize}               
\usepackage{xspace}                
\usepackage[export]{adjustbox}

\usepackage[svgnames]{xcolor}      
\definecolor{bluekeywords}{rgb}{0.13,0.13,1}
\definecolor{greencomments}{rgb}{0,0.5,0}
\definecolor{redstrings}{rgb}{0.9,0,0}

\usepackage{hyperref}  
\hypersetup{
          linkcolor = black,
          citecolor = black }

\usepackage{amsmath}
\usepackage{amsfonts}              
\usepackage{amssymb}               

\usepackage{listings}              
\ifCLASSOPTIONcompsoc
 \usepackage[caption=false,font=normalsize,labelfont=sf,textfont=sf]{subfig}
\else
 \usepackage[caption=false,font=footnotesize]{subfig}
\fi
\usepackage{fixltx2e}

\begin{document}
%
\title{Exploring Computation-Communication Tradeoffs in Camera Systems}


\author{
	\IEEEauthorblockN{Amrita Mazumdar\IEEEauthorrefmark{1},
					  Thierry Moreau\IEEEauthorrefmark{1},
					  Sung Kim\IEEEauthorrefmark{2},
					  Meghan Cowan\IEEEauthorrefmark{1}, \\
					  Armin Alaghi\IEEEauthorrefmark{1},
					  Luis Ceze\IEEEauthorrefmark{1},
					  Mark Oskin\IEEEauthorrefmark{1},
					  and Visvesh Sathe\IEEEauthorrefmark{2}}
	\IEEEauthorblockA{\IEEEauthorrefmark{1}Paul G. Allen School of Computer Science \& Engineering, University of Washington}
	\IEEEauthorblockA{\IEEEauthorrefmark{2}Department of Electrical Engineering, University of Washington}
	\IEEEauthorblockA{\{amrita,moreau,cowanmeg\}@cs.washington.edu, sungk9@uw.edu, \{armin,luisceze,oskin\}@cs.washington.edu, sathe@uw.edu}
					  }

\maketitle

\begin{abstract}

\noindent Cameras are the defacto sensor.
The growing demand for real-time and low-power computer vision, coupled with trends towards high-efficiency heterogeneous systems, has given rise to a wide range of image processing acceleration techniques at the camera node and in the cloud.
In this paper, we characterize two novel camera systems that use acceleration techniques to push the extremes of energy and performance scaling, and explore the computation-communication tradeoffs in their design.
The first case study targets a camera system designed to detect and authenticate individual faces, running solely on energy harvested from RFID readers.
We design a multi-accelerator SoC design operating in the sub-mW range, and evaluate it with real-world workloads to show performance and energy efficiency improvements over a general purpose microprocessor.
The second camera system supports a 16-camera rig processing over 32 Gb/s of data to produce real-time 3D-360$^{\circ}$ virtual reality video.
We design a multi-FPGA processing pipeline that outperforms CPU and GPU configurations by up to 10$\times$ in computation time, producing panoramic stereo video directly from the camera rig at 30 frames per second.
We find that an early data reduction step, either before complex processing or offloading, is the most critical optimization for in-camera systems.

\end{abstract}


\section{Introduction}
\label{sec:intro}
Cameras are the backbone of data processing for applications ranging from social media and entertainment, to surveillance, biomedical devices, and autonomous vehicles.
As these systems continue to specialize and diversify, the traditional interface between camera sensors and general-purpose processors limits optimization for extreme visual computing applications.
Typically, architects employ one of two solutions to enable compute-heavy vision: on-device hardware acceleration, or cloud offload.
Hardware accelerators achieve improved performance and efficiency at the cost of fixed functionality, while offloading data to the cloud relaxes computational constraints at the cost of data communication.
The design tradeoff reduces to balancing computation and communication constraints for a given visual computing workload.
In this paper, we investigate two end-to-end camera systems that push the boundaries of energy efficiency and performance, under these lenses of computation and communication costs.
For each system, we focus on holistically evaluating the full system and  computation-communication tradeoffs across parts of the processing system via ``in-camera processing pipelines.''

The first camera system is an ultra-low-power camera system that recognizes specific users' faces while running on harvested radio frequency (RF) energy.
The ability to run untethered from a power source makes deployment simple, but pushes the design constraints to the extreme end of ultra-low-power design.

The second camera system assembles immersive stereoscopic virtual reality (VR) video processing in real time, requiring significantly more compute and communication performance.
The system consists of 16 4K-resolution cameras, processing hardware, and a network link.
To deliver real-time VR video, the system processes up to 32 Gb/s, making it impractical to transmit the sensor data to a data center for real-time stereo processing and stitching.

\begin{figure}[t]
\centering
    \begin{center}
\includegraphics[width=.48\textwidth]{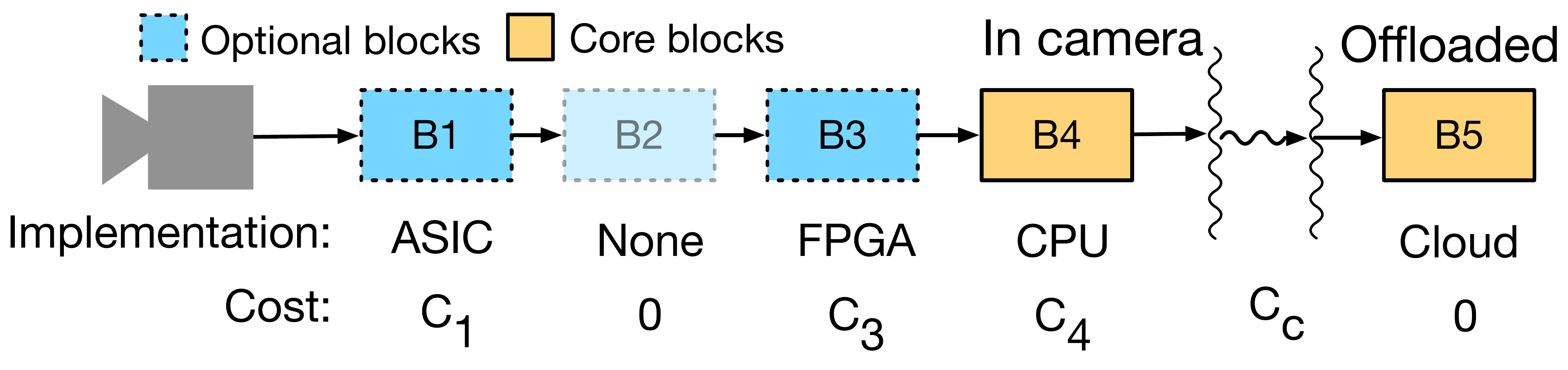}
    \end{center}
    \vspace{-1.5em}
    \caption{Hypothetical in-camera pipeline with opportunities for acceleration. This pipeline uses \textit{core} blocks and some \textit{optional} blocks, and offloads computation to the cloud.}
    \label{fig:generic_pipeline}
    \vspace{-1.75em}
\end{figure}

These systems push the bounds of camera system engineering, at opposing ends of the design space of camera systems---extreme low power, and extreme performance.
Both systems decompose to pipelines of application-specific computation blocks, like the generic in-camera processing pipeline of Figure~\ref{fig:generic_pipeline}.
By characterizing two extreme design points in a common framework, we highlight how these data movement considerations are common across the spectrum of camera applications.
Many other camera systems are likely to exist between the power-constrained and high-bandwidth case studies we investigate.

This paper makes the following contributions:
\begin{itemize}
\item A low-power face authentication accelerator for energy harvesting cameras, with an ASIC evaluation on real-world workloads.
\item A real-time stereoscopic video assembly accelerator for virtual reality, with a CPU, GPU and FPGA comparison.
\item A joint evaluation of computation and communication costs, demonstrating how adding more computation can \textit{reduce} the overall cost of accelerator-based image processing architectures.
\end{itemize}

\section{Background and related work} \label{sec:background}

In-camera processing is not new, and prior work has introduced many in-camera processors~\cite{hauswald2014hybrid}. Our analysis applies in-camera processing pipelines to two highly-constrained applications: low-power face authentication and real-time VR video streaming. In this section, we discuss our general approach to analyzing image processing pipelines and review notable related work in computation offloading, image processing hardware, and similar accelerator designs.

 \textbf{In-camera processing pipelines.}
To characterize in-camera systems in a holistic way, we decompose camera applications into processing pipelines and evaluate the system at the level of functional block, as shown in Figure~\ref{fig:generic_pipeline}.
Considering camera systems at the block granularity helps us gain insight into deciding what processing steps should be included at the camera node, and what implementations (e.g., ASIC, FPGA, GPU) meet an application's requirements.
In the hypothetical pipeline of Figure~\ref{fig:generic_pipeline}, blocks $B_1$, $B_3$, and $B_4$ may be processed in-camera while the output of $B_4$ is offloaded to a central processor such as a multicore or cloud processor.
The block $B_2$ is shown excluded from the pipeline because it does not improve the overall cost.
We define the total cost of the pipeline as the sum of computation costs for in-camera blocks ($C_1$, $C_3$, and $C_4$) and the communication cost ($C_c$) of offloading the output of $B_4$.
We assume the cost of computing in the cloud as ``free'' (relative to computation in the camera) but the cost to get data to the cloud is not (e.g., the camera expends energy to send data).
Hence, one can view the main objective of computing in-camera is to minimize both the data communicated and the computational cost.

In-camera processing pipelines can include \emph{core blocks} essential to the application, and \emph{optional blocks}, which may not directly affect results but can improve efficiency by filtering or pre-processing data.
One optional block is the motion detection block we use in our face authentication pipeline.
While the core block of the pipeline, face authentication, operates on every input frame, an optional motion detection block can reduce the bandwidth and ensuing power consumption of core blocks.
 \textbf{Computation offload.}
Offloading image processing computation from mobile devices to the cloud is well-explored in mobile systems~\cite{pervasive_compute}.
The opposing case for ``onloading'' computation, or keeping computation at the sensor, has grown more popular due to increased image processing demand and privacy concerns~\cite{han-hotos, likamwa-apsys}.
Our approach explores the tradeoff space between offload and onload for two constrained camera systems.

 \textbf{Vision-centric architectures.}
The rise of computer vision and computational photography has inspired a number of computer architectures for efficient image processing.
Flexible vision architectures~\cite{clemons_pmem, hpca_visionsys, myriad15} provide higher performance for image processing and vision applications while maintaining programmability.
Mobile SoCs like Qualcomm's Snapdragon provide image processing functionality for mobile cameras~\cite{snapdragon}.
Vasilyev et al.~\cite{vasilyev_2016} argue towards programmable image processing solutions, but find that custom ASICs are still more energy efficient.
Consequently, we choose to explore fixed-function hardware to meet the constraints of our ultra-low-power or high-performance application targets.

We consider different classes of image processing accelerators for the computational blocks in our case studies; we now detail related work in each class.

 \textbf{In-sensor processing.} Image sensor data is typically captured as an analog signal and converted to a digital signal for processing. Recent work investigated how to improve application efficiency by moving some preliminary processing into the analog domain at the sensor node.
Centeye, for instance, executes analog computation on image sensor signals~\cite{Centeye}.
Other work computed early layers of convolutional NNs at the pixel level~\cite{Chen_2016_ASPVision, redeye}.
Processing can also be performed in the mixed-signal domain~\cite{Alaghi13}.

 \textbf{Face detection accelerators.} We investigate the use of a face detection accelerator as an optional block to filter data in a face authentication pipeline.
Hardware acceleration for the Viola-Jones face detection algorithm has been well-explored for FPGAs and GPUs~\cite{cvpr2007, cho_parallelized_2009,vj_gpu}.
While Bong et al. also present a neural network design using Haar filters as a first step, our work performs a more holistic characterization to optimize the full camera pipeline~\cite{bong20170}.

 \textbf{Neural network accelerators.} NNs have been studied extensively for accomplishing face detection and recognition~\cite{Rowley98,garcia2004convolutional,deepface}.
Researchers are actively working to improve NN performance with custom hardware~\cite{farabet2010, neuflow}.
ShiDianNao~\cite{shidiannao}, specifically, is a CNN accelerator executed in-camera, where the accelerator is placed on the same chip as the image sensor processor, achieving 320mW power consumption.

 \textbf{Depth from stereo accelerators.} Depth from stereo algorithms and their implementations have been well-explored~\cite{scharstein2002taxonomy}.
Stereo vision has been accelerated to real-time with GPUs and FPGAs, but application targets are either very lower resolution or perform badly on defocusing workloads~\cite{pami_gpu_stereo14,date_fpga_stereo14}.

 \textbf{In-camera compression.} Compressing sensor data incurs computation--communication tradeoffs  related to this paper's analyses.
In our VR pipeline, for instance, the output of some blocks might have a better data locality than the previous step, facilitating high compression rates, but lossy compression at the early stages of the pipeline could result in quality degradations.
While we do not explicitly consider compression in our study, compression can be treated as an optional block in in-camera processing pipelines.

\section{Case study: low-power face authentication}
\label{sec:face-auth}

\begin{figure}[b]
\centering
    \begin{center}
\includegraphics[width=.3\textwidth]{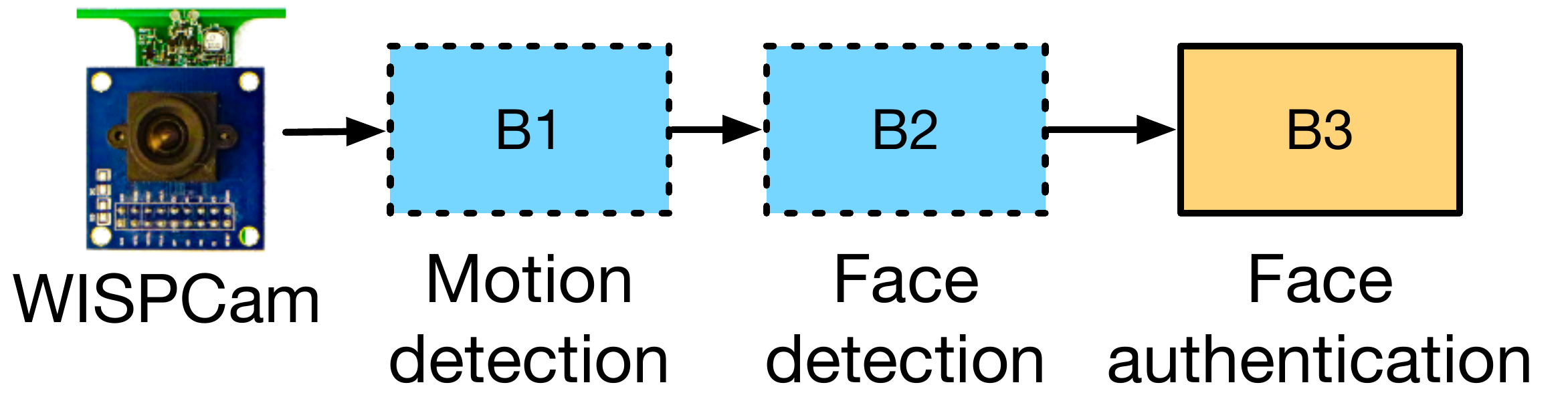}
    \end{center}
        \vspace{-1.5em}
    \caption{Face authentication with battery-free cameras. }
        \vspace{-1em}
    \label{fig:face_recog_pipeline_solution}
\end{figure}

In this section, we characterize a continuous vision pipeline for face authentication based on the WISPCam platform~\cite{wispcam}, a battery-free camera powered by harvested energy.
Face authentication (FA) is a core workload in user-centric continuous mobile vision systems.
In these systems, a camera captures image frames at a continuous frame rate, and an on-node processor performs face recognition on each frame to identify a single user.
We define the core FA function as: given a test face and a reference, decide if the test face matches the reference face.

The WISPCam-based system captures an image at 1 frame per second (FPS) and transmits it over RF, powered by an internal capacitor with harvested RF energy.
We examine how leveraging progressive filtering hardware can dramatically reduce the power consumption of such a system and enable continuous face authentication at low cost.
We construct our FA pipeline around NN-based face authentication, as shown in Figure~\ref{fig:face_recog_pipeline_solution}.
The pipeline has one core block, the NN, and several optional blocks.
We evaluate a low-power NN accelerator design, as well as the benefits of including motion detection and a pre-processing face detection accelerator to reduce input bandwidth to the NN\@.
Because energy efficiency is a primary concern, we design the accelerators to be integrated on-chip with the camera sensor, and processed streaming through the CSI2 camera serial interface.

We first discuss each accelerator design individually, presenting their microarchitectures and the tradeoffs we investigated in each design's algorithm and hardware implementation.
We then evaluate them together on a real-world face authentication workload using real video we collected.

\subsection{Neural network face authentication}

For our face authentication task, we investigate a systolic NN design, based on SNNAP, and explore tradeoffs in neural network (NN) topology, accelerator geometry, and datapath width reduction~\cite{snnap}.
\begin{figure}[tb]
\centering
      \noindent
      \includegraphics[width=0.45\textwidth]{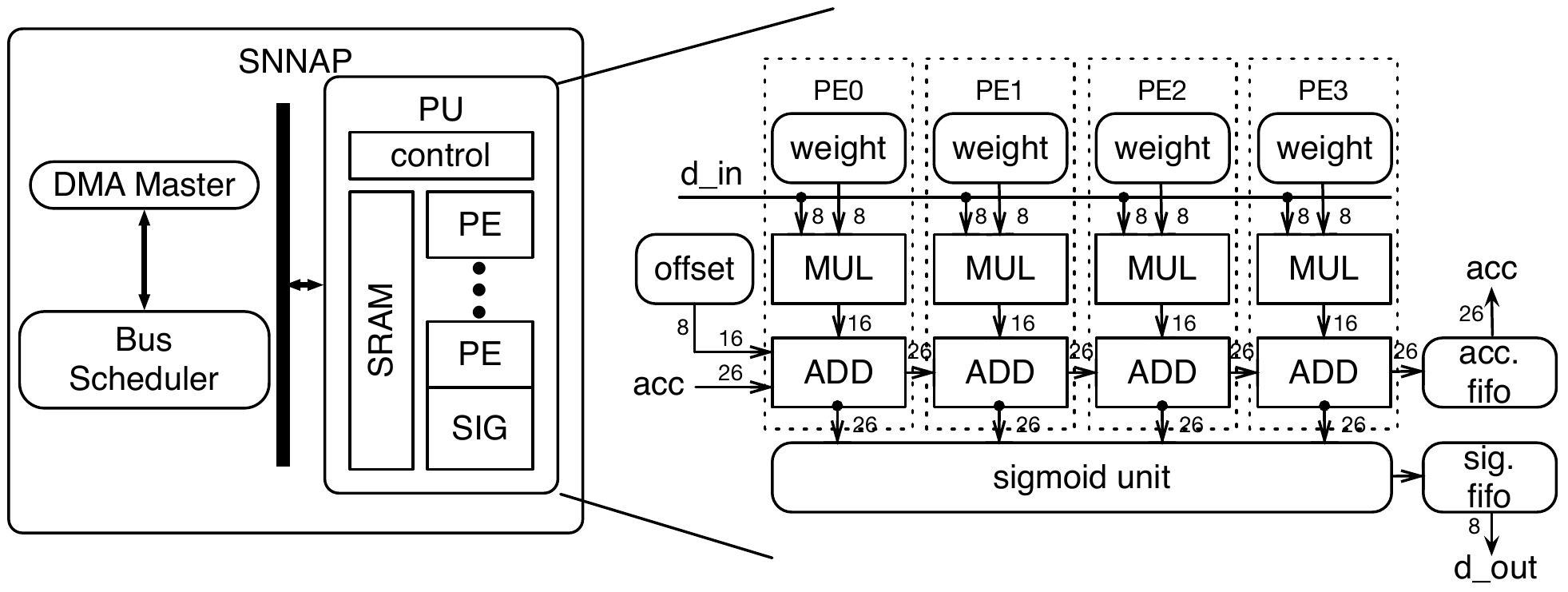}
    \vspace{-1em}
    \caption{NN microarchitecture and processing element details.}
    \vspace{-1.5em}
    \label{fig:snnap-pu}
\end{figure}

\textbf{NN algorithmic tradeoffs.}
We first examine how modifying NN topology affects both classification accuracy and energy dissipation.
We explore the search space by training NNs with Fast Artificial Neural Network Library~\cite{fann} and measuring the achieved accuracy and energy cost.

Increasing the number of layers and neurons directly impacts the memory and computational requirements of the NN\@.
Varying the input size to the NN has a direct impact on performance and accuracy. Using a $5\times5$ low-resolution input window for face detection will lead to a cheap 25-neuron input layer, but results in poor accuracy.
The largest input size our NN supports, $20\times20$ pixels, preserves more details, improving the accuracy of the NN classifier significantly.
This comes at a cost: halving classification error incurs an order-of-magnitude increase in energy.
From this exploration, we select the topologies that give us an optimal accuracy/energy compromise, a $400-8-1$ NN topology with 400 inputs neurons, 8 hidden neurons and 1 output neuron.

To evaluate accuracy tradeoffs, we trained a 400-8-1 NN on 90\% of LFW~\cite{LFWTech}, a popular face recognition benchmark, and tested its accuracy at recognizing a single person's face from the remaining 10\%.
Our evaluation indicates that with a $400-8-1$ topology, we can achieve a 5.9\% classification error overall.
As we discuss on our real-world evaluation, however, our multi-stage approach and real-data workload lowers the true miss rate of 0\%, as the security workload presents many less-challenging lighting and orientation scenarios.


\textbf{NN microarchitecture.}
Our NN microarchitecture uses a single processing unit with multiple processing elements.
Because our face authentication pipeline has wide layers, we found that this design presented enough data parallelism to keep functional unit utilization high for a single processing unit.
Figure~\ref{fig:snnap-pu} shows the datapath of a processing unit composed of four 8-bit processing elements (PEs).
A bus connects the chain of processing elements to a sigmoid unit---a hardware LUT-based approximation of a neuron's activation function.
Each PE has its own weight memory that stores the synaptic weights of the NN locally.
The processing elements perform multiply-add operations in a systolic fashion to evaluate the matrix multiplication that composes each NN hidden and output layer.
A vertically micro-coded sequencer sends commands to each processing element as inputs arrive and outputs are produced to control data movement.


The NN hardware accelerator has a configurable number of PEs, which we use to optimize the geometry of our accelerator.
We fix the frequency and voltage to 30MHz and 0.9V, and explore the design tradeoffs between energy and throughput using post-synthesis physical simulations.
We find an energy-optimal point at 8 PEs: any lower number of PEs introduces scheduling inefficiencies, increasing energy consumption; too many PEs results in underutilized resources and reduced parallelism for the narrow network.

\newbox\mybox
\begin{lrbox}{\mybox}
\begin{lstlisting}[linewidth=.4\columnwidth]
for x in range(0,image_width):
  for y in range(0,image_height):
    faces += classify(x,y,window)
    window *= scale_factor
    if window > image_size:
      return
\end{lstlisting}
\end{lrbox}

\begin{figure}[t]%
    \centering
    \subfloat[Algorithm pseudocode.]{\usebox\mybox \label{fig:vj-algo}}
    \quad
    \subfloat[Cascading classifier.]{{\includegraphics[width=4.5cm,valign=m]{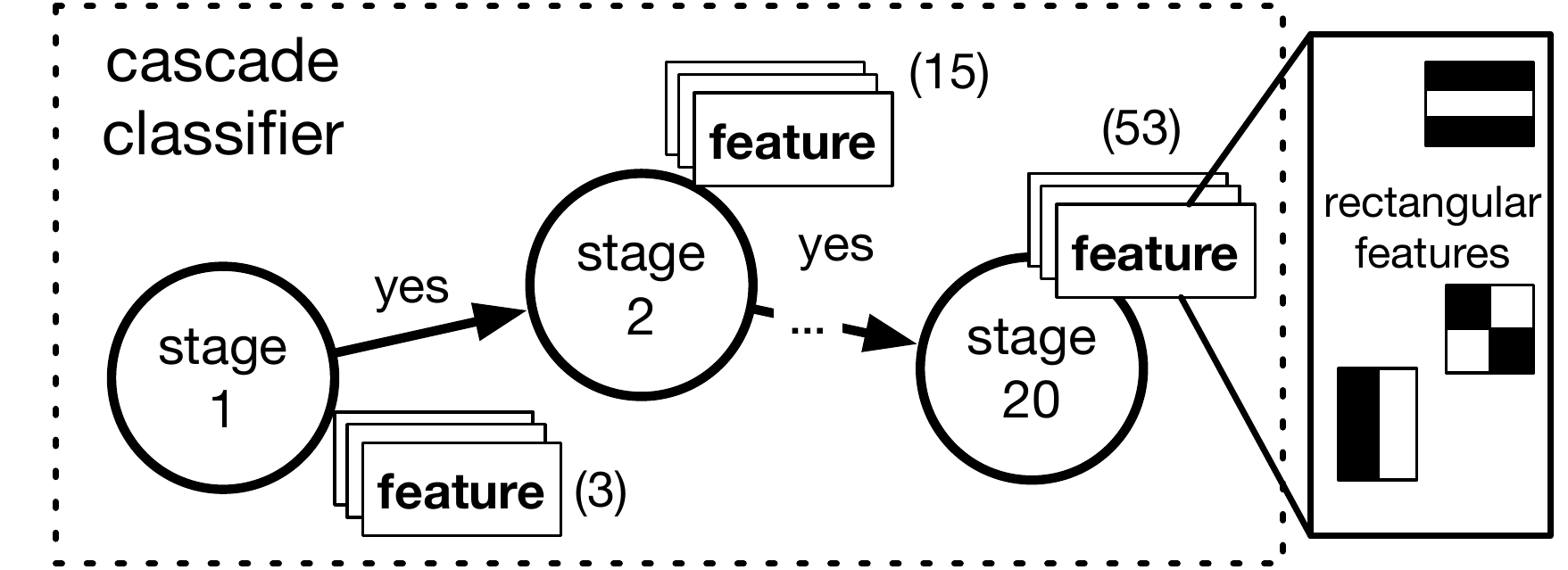} \label{fig:vj-cascade}}}%
    \hfill
    \vspace{-1em}
    \subfloat[Impact of VJ parameters on relative accuracy.]{{\includegraphics[width=0.4\textwidth]{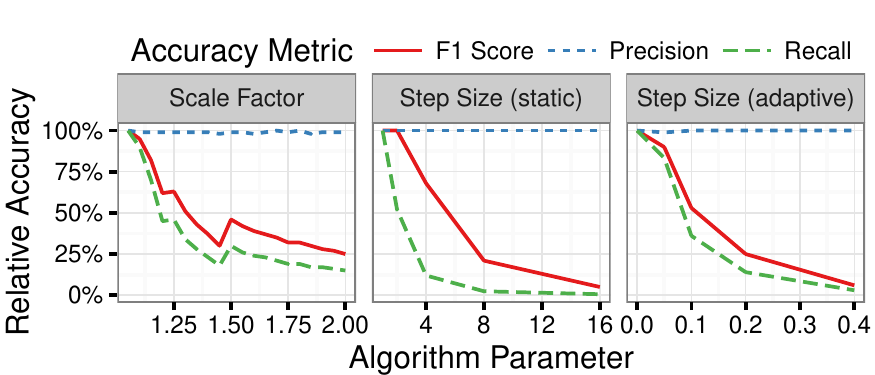}     \label{fig:vj-accuracy}
}}
    \caption{The face detection algorithm slides a window across an image and repeatedly executes a classifier with stages of rectangular features.}%
    \label{fig:vj-stuff}%
            \vspace{-2em}

\end{figure}




\textbf{NN numerical accuracy tradeoffs.}
Power dissipation in the memory and the PEs can be reduced by bit-width reduction.
We used fixed-point functional units and LUT-based approximations of mathematical functions to minimize power and area.
We study the impact of two precision knobs on application accuracy: (1) sigmoid approximation and (2) data bit-width.
We evaluate error as absolute classification accuracy loss relative to a NN implemented with floating-point arithmetic and precise mathematical functions.
We then evaluate fixed-point precision, limiting ourselves to powers of two for memory alignment.

After examining the effect of approximating the sigmoid function with a simple 256-entry look-up table (LUT), we conclude that hardware approximation of the sigmoid function has a negligible effect on accuracy.
For datapath width, both 16-bit and 8-bit implementations of the NN accelerator result in a small 0.4\% accuracy loss relative to a precise floating-point implementation.
The 4-bit datapath however displays a significant accuracy loss on average (over 1\%).
The reduction in datapath width from 16-bit to 8-bit leads to a 41\% power reduction for an 8-PE configuration, so we select 8-bit datapaths as the optimal energy-accuracy point for our NN implementation.

\subsection{In-camera face detection}

The Viola-Jones (VJ) face detection algorithm is a popular computer vision algorithm for fast, accurate face detection ~\cite{vj_journal}. It is widely used in face authentication and other situations where frontal faces are expected and speed is preferred. The algorithm detects faces by scanning a window across the image, evaluating simple rectangular features within the window at each window position. If enough of these features are found at a single window position, then that window is identified as a face. To account for faces of different sizes in an image, the scanning window is scaled and passed over the scene multiple times. The VJ algorithm is well-known because of its simplicity and efficiency, and continues to perform well against more complex algorithms including deformable parts models and convolutional NNs on face detection~\cite{mathias_fd}.

The VJ algorithm is popular specifically because of its high efficiency in non-face windows -- the algorithm optimizes to spend more computation on windows where there is likely to be a face, rather than executing a uniform computation at every window. This optimization is encoded in the \textit{cascade classifier} structure illustrated in Figure~\ref{fig:vj-cascade}, a nested decision tree where progressive levels have increasingly more features to evaluate, and the simple stages must be evaluated positively first before continuing on. The cascading computational style makes VJ a good fit for a pre-filtering accelerator. 

\section{Case study: real-time virtual reality video} \label{sec:vr}

\begin{figure}[t]
\centering
    \begin{center}
\includegraphics[width=.49\textwidth]{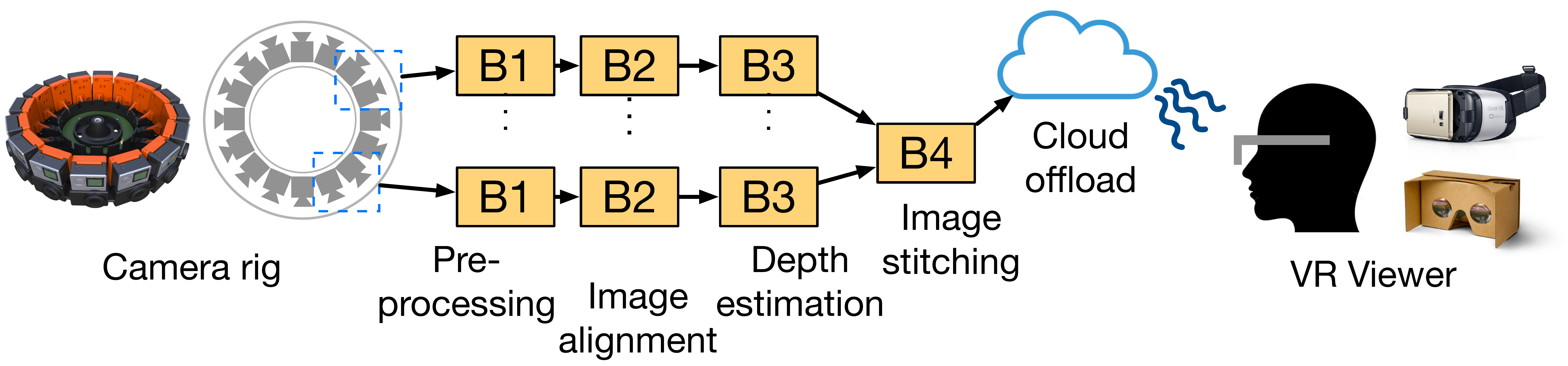}
    \end{center}
    \caption{3D-360$^{\circ}$ virtual reality video generation, capture and viewing devices.}
    \vspace{-1em}
    \label{fig:vr_pipeline_solution}
\end{figure}

\begin{figure}[b]
  \centering
\includegraphics[width=.35\textwidth]{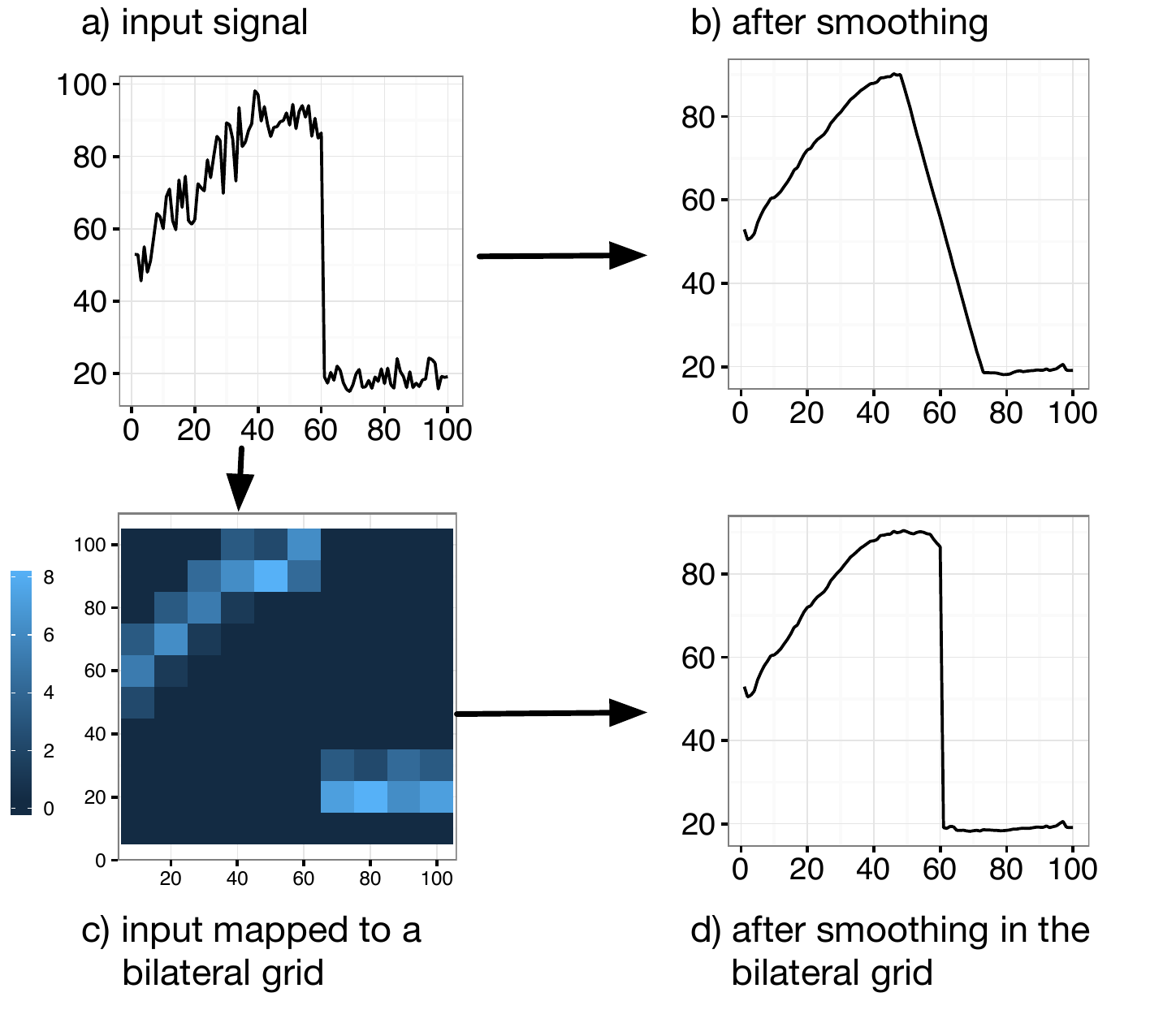}

\caption{The bilateral filter is an edge-aware filter. }
\label{fig:bilat}
\end{figure}

In this section, we investigate the use of in-camera processing for a high-performance, real-time panoramic stereo video rendering application.
As shown in Figure~\ref{fig:vr_pipeline_solution}, the pipeline we consider takes as input the high-resolution camera feeds from a rig of cameras, like Google Jump~\cite{googlejump}, and processes the images into a 360$^{\circ}$ stereo pair viewed on a VR viewer, such as Google Cardboard \cite{googlecardboard}.
The goal is to produce high-quality video streams at a frame rate of 30 frames/sec or more.

Many VR video pipelines pipelines require users to upload camera streams to a cloud service or high-performance computing system---this workflow prevents real-time applications such as live VR video streaming.
While real-time hardware systems for processing VR video are becoming commercially available~\cite{sphere,vahana}, these solutions provide either live panorama processing or stereoscopic 3D, not both.
In our design, we evaluate the performance constraints of this multi-step pipeline and investigate how much in-camera processing is required to achieve real-time VR video generation.
We evaluate how processing at the camera node reduces the bandwidth required for offloading, and how hardware acceleration facilitates a real-time VR system.

Camera rigs for recording stereoscopic panorama videos capture a multi-camera scene and compute a depth map for each pair of cameras in the rig. These depth maps are composited together from multiple pairwise-camera pipelines into a single 3D-360$^{\circ}$ video.
For our application, we seek to meet a real-time frame rate of 30 frames/sec, so we optimize our design for the cost of throughput.
We define the communication cost as the bandwidth in and out of each block. Since all the pipeline blocks and offloading can be pipelined, the slowest step will dominate overall throughput.
Among the blocks shown in Figure~\ref{fig:vr_pipeline_solution}, the depth estimation step has the lowest bandwidth and throughput.
In this section, we describe the depth estimation algorithm used for this block, how we map the algorithm to a high-throughput accelerator, and evaluate the system's computation-communication tradeoffs towards real-time results.

\begin{figure}[t]
\centering
    \begin{center}
\includegraphics[width=.45\textwidth]{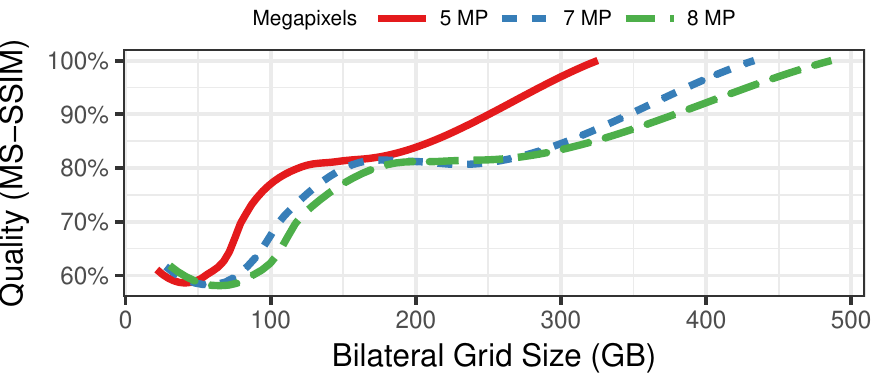}
    \end{center}
    \vspace{-1em}
    \caption{Using a smaller bilateral grid is cheaper to compute but degrades the quality of the output depth map, even at high image resolutions. }
    \label{fig:vr-res-qual}
\end{figure}

\begin{figure}[b]
  \vspace{-1em}

\centering
    \begin{center}
\includegraphics[width=.4\textwidth]{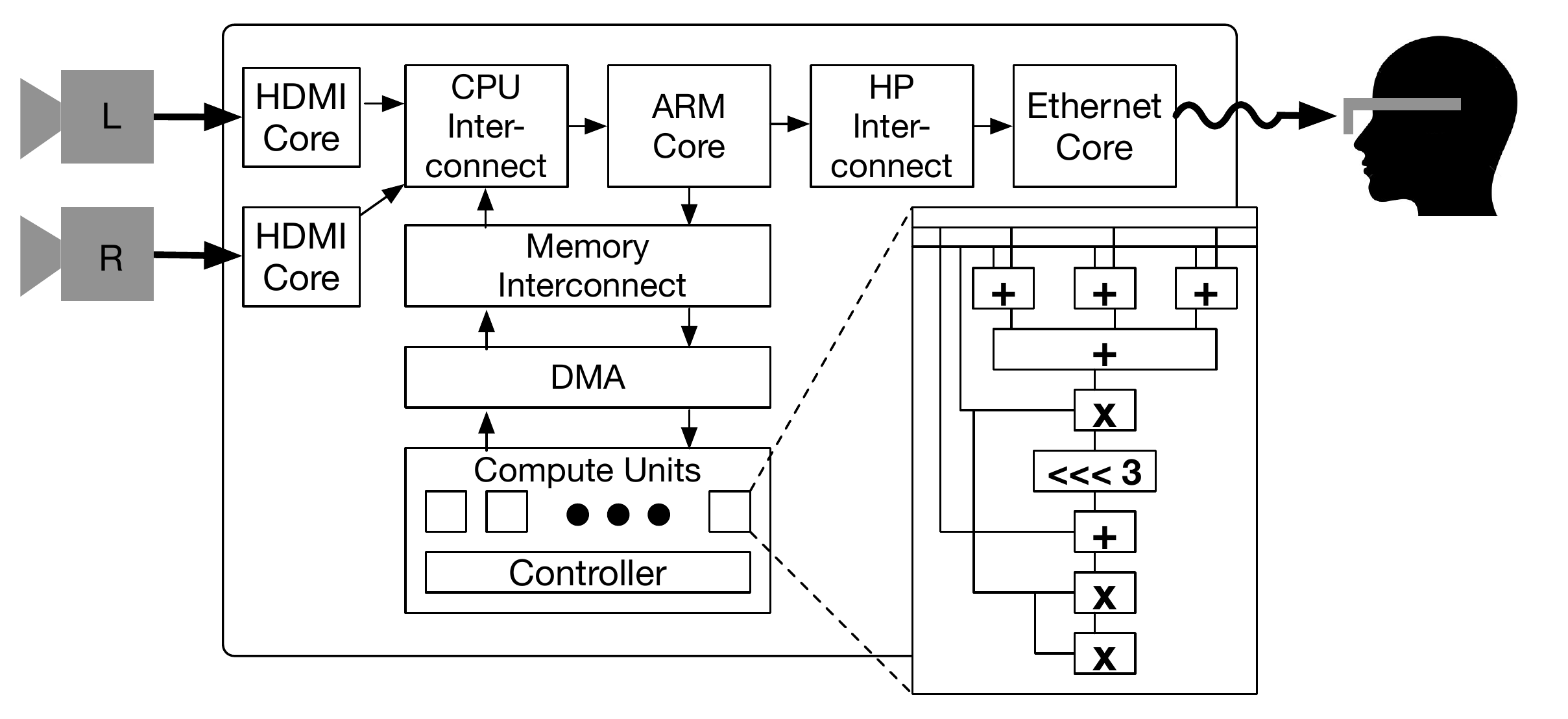}
    \end{center}
    \vspace{-1em}
    \caption{VR accelerator architecture on a Xilinx Zynq SoC.}
    \label{fig:vr-arch}
\end{figure}

\subsection{Depth maps from bilateral-space stereo}

We base our design for fast and accurate stereo processing on the state-of-the-art bilateral-space stereo algorithm (BSSA)~\cite{barron2015}. Typically, global stereo algorithms generate a depth map from a pair of images by computing a rough disparity, or difference in space, between pixels, and then refining that disparity until a cost function has been minimized~\cite{scharstein2002taxonomy}. Instead of computing disparities per-pixel, BSSA resamples the problem into a different representation, \emph{bilateral-space}, before computing the disparity. In the bilateral domain, simple local filters are equivalent to costly, global edge-aware filters in pixel-space---consequently, disparity refinement is much faster in bilateral space. We perform BSSA in a \emph{bilateral grid} data structure, where pixels are mapped to a grid vertex, or bin, in bilateral-space. Filtering in the bilateral grid results in faster, higher-quality output than comparable techniques~\cite{barron2015}.

We illustrate the operation of a bilateral filter in Figure~\ref{fig:bilat}. For simplicity, we demonstrate a 1D signal, instead of a 2D image signal. Our stereo algorithm seeks to smooth the noisy signal of Figure~\ref{fig:bilat}a, which has a sharp edge. Applying a 1D moving average on Figure~\ref{fig:bilat}a results in Figure~\ref{fig:bilat}b, which has less noise and a smoothed-out edge. A bilateral filter performs the same smoothing operation while preserving the edge of Figure~\ref{fig:bilat}a. The signal is mapped to bilateral space as in Figure~\ref{fig:bilat}c, where neighboring pixels with significantly different intensity values will have a large distance in 2D-space. Smoothing this signal in the bilateral domain with a 2D moving average allows the signal to maintain edges. Figure~\ref{fig:bilat}d shows the result after filtering in bilateral-space.


Instead of a simple filter like moving average, BSSA maps a noisy depth map to a bilateral grid, refines the depth map by solving an optimization problem, and remaps the bilateral-grid result to pixel-space. Varying the number of pixels that map to a grid vertex impacts the time to compute the stereo refinement for a frame, and also the quality of the depth map. Figure~\ref{fig:vr-res-qual} demonstrates the tradeoff between stereo image quality and bilateral grid size to be processed for high-resolution input images. Here, we scaled bilateral grid sizes from 4 pixels-per-grid-vertex to 64 in each of three dimensions in a bilateral grid and evaluated the resulting impact on quality using MS-SSIM~\cite{msssim}. We find the resolution of the input images is less impactful than choosing an appropriate grid size to balance quality and computational complexity.

\begin{figure}[tb]
\centering
    \begin{center}
\includegraphics[width=.48\textwidth]{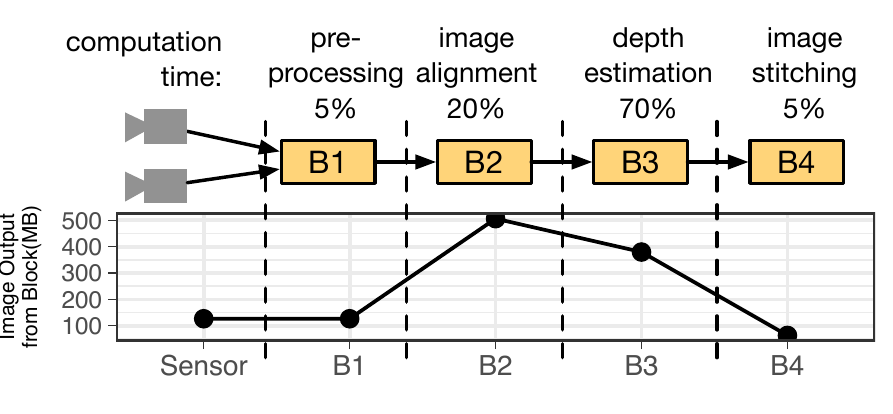}
    \end{center}
    \vspace{-1em}
    \caption{Computation distribution and output data size for blocks in a VR video pipeline (2 of 16 cameras). }
    \label{fig:vr-data-scale}
    \vspace{-1em}
\end{figure}

\subsection{BSSA accelerator design on FPGAs}

\begin{figure*}[t!]
\centering
    \begin{center}
\includegraphics[width=1.0\textwidth]{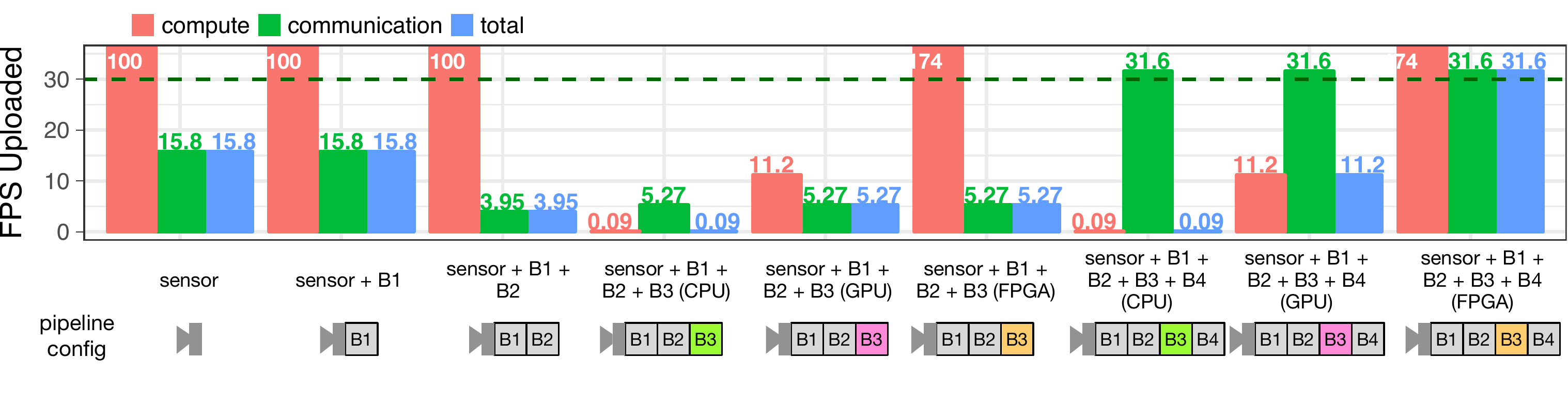}
    \end{center}
    \vspace{-2.5em}
    \caption{Pipeline configurations with different bilateral smoothing implementations (CPU, GPU, FPGA), and resulting upload rates (frames per second). Only the full pipeline with FPGA acceleration can meet a 30 FPS upload requirement.}
    \label{fig:vr-fps}
    \vspace{-1em}
\end{figure*}

We design and implement our processing flow in Verilog on the Xilinx Zynq-7020 SoC~\cite{ZYNQ}. Figure~\ref{fig:vr-arch} depicts the high-level architecture of our system. We implement the initial full pipeline in software to run on the Zynq's CPU, and then design an AXI-Stream-compliant FPGA accelerator for depth refinement that can be invoked by the software. The CPU prepares the bilateral-grid data structure with pixels mapped to grid vertices, and transfers them via DMA to the FPGA fabric. The hardware accelerator processes the vertices with the bilateral-space filtering and streams them back to the CPU, where the bilateral-grid-filtered result is converted into the fully-processed depth map.

Figure~\ref{fig:vr-data-scale} shows the processing break-down for our pipeline in time consumption and the image data size produced by each block. We find that the depth estimation block, $B_3$, consumes the greatest computation time as well as the largest amount of data, from $B_2$. We thus focus on applying FPGA acceleration to this block, and then evaluate the impact of accelerating this block on pipeline throughput.

Applying the computation of $B_3$ to a high-resolution video is equivalent to applying millions of blurs to the bilateral grid representation of the video frames. Across a single frame, most of these filters can run in parallel, so we designed streaming compute units to run bilateral filters on a stream of grid vertices. We find that BSSA requires at least 32-bit floating-point precision to produce high-quality depth maps, and use DSP units on the FPGA fabric to compute efficient floating-point operations. Each compute unit requires 18 DSP units in our design, so we can scale up to 12 parallel compute units on the ZC702. However, we project that if we scale up to a top-of-the-line Xilinx Virtex UltraScale+ FPGA, we can parallelize up to 682 compute units, which are more than enough for real-time operation. Table~\ref{table:vr-resources} summarizes the setup we use in our evaluation and resource requirements for real-time performance with a 16-camera system.

\subsection{Evaluation}

\begin{table}[t!]
  \centering
  \vspace{-.25em}
\caption{Requirements for FPGA acceleration platform.}
  \vspace{-.5em}
  \resizebox{\columnwidth}{!}{%
  \begin{tabular}{ l @{\hskip 6pt}l @{\hskip 6pt}l @{\hskip 6pt}l  }
\toprule
 & \textbf{Resource} & \textbf{Evaluation} & \textbf{Target} \\
\midrule
System   & FPGA Model       & Zynq-7000 & Virtex UltraScale+ \\
         & FPGA (\#)        & 1                 & 16                       \\
         & Cameras          & 2                 & 16                       \\
\midrule
Per FPGA & Logic            & 45.91\%           & 67.10\%                  \\
         & RAM              & 6.70\%            & 17.60\%                  \\
         & DSP              & 94.09\%           & 99.98\%                  \\
         & Clock (MHz)      & 125               & 125       \\
\bottomrule
\end{tabular}
}
    \vspace{-1em}
\label{table:vr-resources}
\end{table}

\textbf{Experimental setup.} We compare our FPGA results on the Zynq platform to CPU and GPU baselines. The Zynq includes a Dual ARM Cortex-A9 and a Xilinx FPGA, all fabricated at TSMC 28nm technology. We implement the CPU baseline on the Zynq's Dual ARM Cortex-A9 as a proxy for a mobile-grade CPU, and evaluate the GPU on an NVIDIA Quadro K2200. Both baselines execute optimized BSSA code written and tuned with Halide~\cite{halide}.

\textbf{Methodology.} We consider the throughput of the data output as the ``communication cost'' for offloading, and the cost to compute the pipeline block as the ``computation cost''. We treat the communication cost as fixed for each block; it is simply the cost of offloading the data from each block, as shown in Figure~\ref{fig:vr-data-scale}. For all blocks except disparity refinement, we assume the computation cost to be the compute time evaluated using the ARM CPU baseline's performance numbers. We average the compute time for the disparity refinement block over five executions of the kernel over a frame. Because this processing flow can be pipelined across frames in a video stream, the ``total cost'' of the system can be considered to be dominated by the lowest-throughput block of the system.

\textbf{Computation-communication tradeoffs.} Figure~\ref{fig:vr-fps} shows the runtime results of different pipeline configurations, uploaded on a networked connection to a viewing device supporting at least 30 FPS. We seek to uncover scenarios in which both computation and communication surpass our minimum frame rate of 30 FPS---if one or both costs falls below the threshold, the system cannot support real-time operation.

For the first three scenarios, the cost of doing little computation before offloading is cheap, even on the ARM core, but the communication cost for the raw captured data falls short of our 30 FPS threshold. Computing the disparity refinement in $B_3$ is more costly, and the CPU and GPU implementations are not fast enough to support real-time operation. Moreover, the cost of offloading the computed depth maps before image stitching is significantly lower.

The computation cost of image stitching in $B_4$ is marginal compared to BSSA, as well, and the resulting FPS is virtually the same. The data size to communicate after $B_4$, however, is much smaller, as illustrated in Figure~\ref{fig:vr-data-scale}, and is the only data size small enough to support real-time uploading. We find that the configuration with all the blocks processed in-camera and $B_3$ mapped to the FPGA is the only configuration where both computation and communication pass the threshold and support real-time processing.

Our analysis indicates that this camera system is primarily constrained by network bandwidth. For our evaluation, we assumed transfer speeds of 25 Gigabit Ethernet. As network connections grow faster, our results will trend towards offloading computation right off the sensor. For instance, at a hypothetical ultra-high-throughput network link of 400-Gb Ethernet, the 16-camera output can be uploaded at 395 FPS, reducing the efficiency incentive for in-camera processing in this scenario.

\vspace{-.5em}
\section{Conclusions}
 \label{sec:conclusion}

Cameras have become the dominant sensor in mobile systems, and complex image processing pipelines are now standard.
In this paper, we use the notion of ``in-camera processing pipelines'' to thoroughly characterize the design of two camera systems at the extreme ends of the energy and performance scaling limits of current hardware.
Our face authentication camera system, for instance, runs entirely on harvested energy, pushing the limits of ultra-low power computation.
Our virtual reality camera system requires significantly more in-camera processing and data communication resources than traditional imaging platforms.
Our results highlight how design parameters for individual accelerators can influence the full-system execution behavior, as well as shape decisions about whether to process a computation block in the camera or offload the computation.

We characterize in detail how even the most power-efficient neural network design performs significantly better when adding computation earlier in the pipeline to effectively filter the image data.
Our VR pipeline highlights how computational stages that expand the data size are inefficient in isolation, and can be better optimized in concert with their down-stream components.

Power and performance constraints require increasingly efficient computational platforms, and architects will continue to look to hardware acceleration to enable challenging applications.
As we demonstrate in this paper, even tightly-optimized accelerators can fail to improve performance if they fail to consider full-system communication challenges. 
Given the growth of image data production and requirements of modern vision and graphics algorithms, future applications require a full system approach to maintain power and performance efficiency of camera designs.

\section{Acknowledgements}

We thank members of the Sampa lab and the anonymous reviewers for their feedback on earlier versions of this work. This work was supported in part by the National Science Foundation under Grant CCF-1518703, by C-FAR, one of the six SRC STARnet Centers, sponsored by MARCO and DARPA, and a generous gift from Google.

\bibliographystyle{IEEEtranS}
\bibliography{paper}

\end{document}